\documentclass[letterpaper]{article}

\usepackage[T1]{fontenc}

\usepackage{geometry}
\usepackage{url}
\usepackage{xcolor}  
\usepackage{setspace}

\usepackage{amsmath,amssymb,amsfonts}
\usepackage{bm}  

\usepackage{achemso}

\usepackage{graphicx}
\usepackage{float}
\newfloat{scheme}{htbp}{los}
\floatname{scheme}{Scheme}
\floatname{chart}{Chart}
\newfloat{graph}{htbp}{loh}

\usepackage{chemformula} 
\usepackage[version = 4]{mhchem} 

\usepackage{authblk}
\author[1]{Xingran Xu$^{*,\dagger,}$}
\author[1]{Yong-Jia Chong$^{*,}$}
\author[1]{Zhi-Hao Sun}
\affil[1]{School of Optoelectronic Information and Physical Science, Jiangnan University, Wuxi,214122 ,China}

\title{Edge of Stability in Nonlinear Photonic Reservoirs: Universal Design Principle for Exciton-Polariton Computing}
\date{* These authors contribute equally. \\
$\dagger$ Email: thoexxr@hotmail.com}

\begin{document}

\maketitle

\begin{abstract}
Reservoir computing has emerged as a powerful paradigm for harnessing the intrinsic dynamics of physical systems to perform complex information processing. Here, we theoretically and numerically investigate a reservoir computing system based on the two-dimensional discrete complex Ginzburg-Landau equation, a fundamental model describing driven-dissipative exciton-polariton lattices. We apply this framework to the challenging 52-class handwritten letter recognition task, decomposing temporal input signals to simulate word recognition. Our system achieves a high test accuracy, significantly outperforming a linear baseline. Through systematic parameter scans and nonlinear dynamics analysis, we establish a quantitative link between computational performance and the underlying physics: optimal classification occurs near the edge of stability, where the maximum Lyapunov exponent approaches zero. This principle holds universally across different control parameters, including nonlinear coupling and linear gain. Furthermore, we validate the practical feasibility of this approach by deploying the trained model on a Raspberry Pi 5 edge device. These results not only demonstrate the potential of polariton lattices for neuromorphic computing but also provide a general design principle for optimizing physical reservoir computing hardware.
\end{abstract}

\section*{Keywords}
exciton-polariton; reservoir computing; neuromorphoc computing;  



\section{Introduction}

Reservoir computing (RC), a bio-inspired framework for temporal information processing also known as echo state networks or liquid state machines, has emerged as a powerful paradigm for harnessing the dynamics of physical systems to perform complex computations~\cite{Jaeger2004,Maass2002,Lukosevicius2009,Verstraeten2005}.
In this framework, a fixed, high-dimensional dynamical system—the reservoir—maps input signals into a feature space, where only a simple linear readout layer is trained via regression. This architectural separation drastically reduces training complexity to a convex optimization problem, making RC particularly attractive for physical hardware implementations where fine-grained control over individual connections is often infeasible.

A wide range of physical platforms have been explored for RC hardware, including optoelectronic delay lines~\cite{Larger2017,Paquot2012}, all-optical integrated circuits~\cite{Vandoorne2014}, memristor arrays~\cite{Du2017}, spintronic oscillators~\cite{Torrejon2017}, mechanical oscillators~\cite{Coulombe2017}, and even fluidic systems~\cite{Fernando2003}; for comprehensive reviews, see Tanaka~\textit{et~al.}~\cite{Tanaka2019}. Among these, photonic systems are particularly attractive due to their potential for ultrafast processing rates and parallel operation~\cite{Brunner2013,Appeltant2011}, with recent advances demonstrating coherent nanophotonic circuits for deep learning~\cite{Shen2017}. Beyond classical platforms, quantum reservoir computing has been proposed using disordered-ensemble quantum dynamics~\cite{Fujii2017}, and large-scale neuromorphic chips such as TrueNorth have demonstrated million-neuron spiking networks~\cite{Merolla2014}.

Recent work \cite{Opala2019} proposed a RC scheme based on the complex Ginzburg--Landau equation (CGLE), one of the fundamental models of nonlinear wave phenomena~\cite{Aranson2002}. The CGLE provides a universal description of weakly nonlinear spatiotemporal systems with gauge invariance \cite{Aranson2002}, spanning applications from hydrodynamics and chemical reactions \cite{Vanag2001} to superconductors \cite{Dorsey1992}, ultracold gases \cite{Madison2000}, and semiconductor lasers \cite{Hohenberg1994}. The CGLE framework can also encompass exciton-polariton microcavity systems, where non-equilibrium exciton-polaritons emerge as hybrid light-matter quasiparticles from the strong coupling between excitons and photons confined within semiconductor microcavities \cite{Weisbuch1992,Kasprzak2006}. Benefiting from this unique nature, they inherit the exceptionally low effective mass and high propagation velocity of photons, alongside the strong nonlinear interactions and environmental sensitivity of excitons \cite{Deng2010,Kavokin2007,Byrnes2014}. These distinctive characteristics position exciton-polaritons as a versatile platform for a wide range of photonic applications, including low-threshold coherent light sources, non-Hermitian skin effect\cite{Xu2021,XU2022,Xu_2025,Bao2023,Liang_2025}, all-optical logic gates \cite{Li_2024,Feng_2021}, single-photon emitters \cite{Do2024}, and highly efficient light-emitting devices for integrated on-chip photonics \cite{Mischok_2023,Hu_2024}.

In the realm of quantum computing and information processing \cite{ValleInclanRedondo2024,Xu2025}, exciton-polaritons offer several profound advantages. First, as composite bosons, they can undergo Bose-Einstein condensation under non-equilibrium conditions, forming macroscopic quantum states that serve as ideal platforms for simulating complex many-body quantum systems and topological phenomena \cite{Camacho_Guardian_2020,Wang_2017}. Second, their intrinsic strong optical nonlinearity and ultrafast picosecond-scale response \cite{Walker2015,Carusotto2013} enable high-speed all-optical signal processing, promising to overcome the speed bottlenecks of conventional electronics and achieve data processing rates on the order of Tbit/s \cite{Amo2009,Blaicher_2020}. Furthermore, the advent of wide-bandgap materials such as gallium nitride (GaN) and perovskites has enabled stable exciton-polariton condensation at room temperature, overcoming the stringent cryogenic requirements of early systems \cite{Su2020b,K_dziora_2024}. This capability to sustain robust quantum effects at ambient conditions significantly lowers the operational barriers for quantum devices, paving the way for the development of low-power, highly scalable on-chip quantum simulators and quantum information processors.  

Polariton reservoirs have already delivered impressive results, from accurate classification of handwritten digits \cite{Opala2019,Gan_2025} to precise reconstruction of quantum states \cite{ghosh2021reconstructing}. The integration of quantum neuromorphic principles with reservoir network architectures and Fourier neural operator methodologies is now propelling the field toward practical, scalable all-optical computing platforms that promise real-time artificial intelligence inference. However, its classification capability was limited to 10-class problems on a small lattice. Several important questions remain open. First, how does the system scale to classification tasks with a substantially larger number of classes, such as the 52-class EMNIST letter recognition benchmark~\cite{Cohen2017} where case-sensitive confusions pose an additional challenge? Second, and more fundamentally, what is the precise relationship between the nonlinear dynamical properties of the CGLE reservoir---characterized by quantities such as the maximum Lyapunov exponent (MLE), effective rank, and temporal correlation structure---and its classification performance? Understanding this relationship is essential for principled optimization of physical reservoir computers beyond empirical parameter tuning.

In this work, we address these questions through a systematic study of CGLE lattice reservoir computing applied to the EMNIST 52-class handwritten letter recognition task~\cite{Cohen2017}.With a $30\times30$ lattice (900~nodes), we classify 26 uppercase and 26 lowercase letters (A--Z, a--z), a significantly more challenging problem than 10-class digit recognition due to the deliberate case-sensitive confusion built into EMNIST.Our aim is not to compete with state-of-the-art deep convolutional networks, which achieve $>$85\% accuracy on this benchmark~\cite{Cohen2017,LeCun2015}, but to use the 52-class task as a challenging testbed for characterizing how the nonlinear dynamics of a physical reservoir govern its computational performance. We perform comprehensive parameter scans across five physical control parameters and characterize the nonlinear dynamics of the reservoir through the MLE, effective rank of the reservoir state matrix, and autocorrelation function across a wide range of both the nonlinear coupling strength $g$ and the linear gain $\gamma$.We also visualize the reservoir activation patterns for individual letters and demonstrate edge deployment on a Raspberry Pi~5 \cite{RaspberryPi5}. Our results reveal that optimal classification performance consistently occurs near the stability boundary where the MLE approaches zero, establishing a quantitative connection between nonlinear dynamics and computational capability in CGLE reservoir computers.

\section{Model}

\subsection{Complex Ginzburg--Landau lattice reservoir}
Single-step word recognition poses a significant challenge due to the limited number of physical nodes available in polariton systems. To address this constraint, we decompose words into individual letters, as illustrated in Fig. \ref{fig:schematic}(a). Each letter is subsequently processed by the exciton-polariton reservoir computing system, as shown in Fig. \ref{fig:schematic}(b).The exciton-polariton system receives the letter image, which is then transformed within a DBR-semiconductor-DBR sandwich-like structure for neuromorphic computing. The system can be considered by a CGLE on a two-dimensional square lattice of weakly coupled nodes, which describes the dynamics of the complex node amplitude $\psi_n$ \cite{Opala2019,Opala_2023,Kavokin_2022}:

\begin{equation}
\label{eq:CGLE}
\frac{d\psi_n}{dt} = P \sum_{m} W_{nm}^{\rm in} u_m - i\sum_{m \in \mathcal{N}(n)} W_{nm} \psi_m
+ \left(\gamma - \Gamma|\psi_n|^2 - ig|\psi_n|^2\right)\psi_n,
\end{equation}

\noindent where the first term on the right-hand side represents coherent signal injection.
The scalar $P$ is the overall input scaling factor; the random matrix elements $W_{nm}^{\rm in}$ are drawn independently from $\mathcal{U}[-1/2, 1/2]$, such that the effective injection weights $P \cdot W_{nm}^{\rm in}$ are uniformly distributed over $[-P/2, P/2]$.
The second term describes nearest-neighbor and next-nearest-neighbor coupling within the lattice, where $\mathcal{N}(n)$ denotes the set of coupled neighbors of node $n$.
The third term contains the local nonlinear dynamics: $\gamma$ is the effective linear gain, $\Gamma$ is the nonlinear damping coefficient that saturates the amplitude, and $g$ is the conservative nonlinear coefficient governing phase-amplitude coupling.

The coupling matrix $W_{nm}$ is real, symmetric, and random, with nearest-neighbor and next-nearest-neighbor (diagonal) couplings drawn from uniform distributions, with the latter assigned weaker strengths to reflect distance-dependent coupling in photonic systems such as microcavity arrays or waveguide lattices.
We note that the factor $-i$ multiplying the coupling sum in Eq.~(\ref{eq:CGLE}) renders the inter-node coupling purely reactive (coherent phase rotation), in contrast to dissipative (amplitude-mixing) coupling commonly found in other RC architectures.
The input weight matrix $W_{nm}^{\rm in}$ is a real-valued random $N^2 \times D$ matrix whose entries control the injection strength, where $D$ is the input dimension.

The baseline physical parameters place the reservoir in a stable regime near the threshold of the trivial solution $\psi_n = 0$, providing a reasonable starting point for subsequent optimization; specific values are given in the following context.

 We propose a physical implementation scheme for the CGLE reservoir based on a GaAs/AlGaAs exciton-polariton microcavity platform, as illustrated in Fig.~\ref{fig:schematic}(c). (The numerical results reported in this work are obtained via Euler integration of Eq.~(\ref{eq:CGLE}) with time step $dt = 0.03$, i.e., 224~steps per letter.)

The platform employs a GaAs/AlGaAs semiconductor microcavity with embedded InGaAs quantum wells---the standard exciton-polariton system, where room-temperature condensation has been demonstrated in both inorganic and organic systems \cite{Su2020b,K_dziora_2024}. Strong coupling between quantum well excitons and cavity photons (Rabi splitting $\sim$5--15~meV) produces exciton-polaritons at each lattice site. The $N\times N = 900$-node lattice is defined optically by patterning a non-resonant pump laser with a spatial light modulator (SLM), eliminating the need for physical etching of the cavity.

The mapping between Eq.~(\ref{eq:CGLE}) parameters and physical quantities is as follows. The linear coefficient $\gamma$ represents the net gain---the difference between the pump-induced gain and the intrinsic cavity photon loss rate $\alpha = \gamma_c \sim 0.1$--$1\;\text{ps}^{-1}$, the latter set by the DBR reflectivity (20--30 AlAs/GaAs layer pairs, $\sim$99\% reflectivity). The pump power provides the most direct experimental control; optimal classification in our study occurs at $\gamma = -0.10$, corresponding to operation slightly below the net-gain threshold where cavity loss marginally exceeds pump gain. The nonlinear damping $\Gamma$ arises from exciton--exciton scattering at high polariton densities and is largely material-intrinsic ($\sim$0.1--0.3 in dimensionless units). The conservative nonlinear coefficient $g$ corresponds to the Kerr nonlinearity originating from exciton--photon detuning $\delta = E_{\rm cav} - E_{\rm exc}$; it can be tuned in both magnitude and sign by adjusting the cavity length (changing the photon energy) or the temperature (changing the exciton energy), with $\delta > 0$ yielding $g > 0$. The inter-node coupling $W_{nm}$ originates from photon propagation within the cavity plane via TE--TM splitting; the coupling strength is controlled by the lattice spacing ($\sim$5--20~$\mu$m between adjacent sites, yielding dimensionless nearest-neighbor couplings $\sim$0.1--0.5 and next-nearest-neighbor $\sim$0.05--0.25), which can be adjusted by reprogramming the SLM pattern. The input scaling $P$ corresponds to the pump modulation depth, and the input weight matrix $\mathbf{W}^{\rm in}$ is implemented as a static amplitude mask on the SLM, with element values drawn from $\mathcal{U}[-P/2, P/2]$ and held fixed throughout both training and inference.

 The input signal $\mathbf{u}(t)\in\mathbb{R}^{28}$, obtained by row-wise scanning of a $28\times28$ handwritten letter image at $\tau_{\rm pixel} = 8$ time steps per row (total duration $224\cdot dt$), is encoded as the intensity modulation of each pump spot. In a transmission geometry, the polariton emission leaking through the bottom DBR is imaged directly onto a CCD or CMOS sensor at the end of the input sequence, yielding the intensity $I_n = |\psi_n(t_E)|^2$ at each of the 900~lattice sites---the readout modality demonstrated in the first polariton reservoir computing experiment \cite{Opala2019}. Since the present classification scheme uses only the squared modulus as the feature vector, phase retrieval is not required for the core pipeline; however, should full complex-valued state information become desirable (e.g., for enhanced feature dimensionality), the phase $\angle\psi_n$ could be recovered via off-axis digital holography using a reference beam split from the pump laser, a technique successfully applied to polariton condensates for phase-resolved imaging. After optical readout, the 900-dimensional feature vector $\mathbf{I} = (I_1,\ldots,I_{900})$ is processed digitally by the pre-trained linear readout layer $\mathbf{W}^{\rm out}$ (a $52\times900$ matrix, 46,800~parameters trained offline via $L_2$-regularized multinomial logistic regression with the L-BFGS algorithm), followed by a 52-class softmax to produce the recognized letter.

With an intrinsic polariton response time of $\sim$10--100~fs, the theoretical classification throughput exceeds $10^{11}$ letters per second. Practical operation, limited by the SLM refresh rate ($\sim$1--10~kHz) and camera frame rate ($\sim$1--100~kHz), would still achieve an estimated $10^4$--$10^6$ letters per second---comparable to existing photonic reservoir computers \cite{Larger2017,Vandoorne2014}. With continuing advances in wide-bandgap materials such as GaN and perovskites enabling room-temperature polariton condensation \cite{Su2020b,K_dziora_2024}, the experimental realization of this scheme is an increasingly near-term prospect.

\subsection{Input encoding and readout}
\begin{figure}[htbp]

	\centering
	\includegraphics[width=1.0\textwidth]{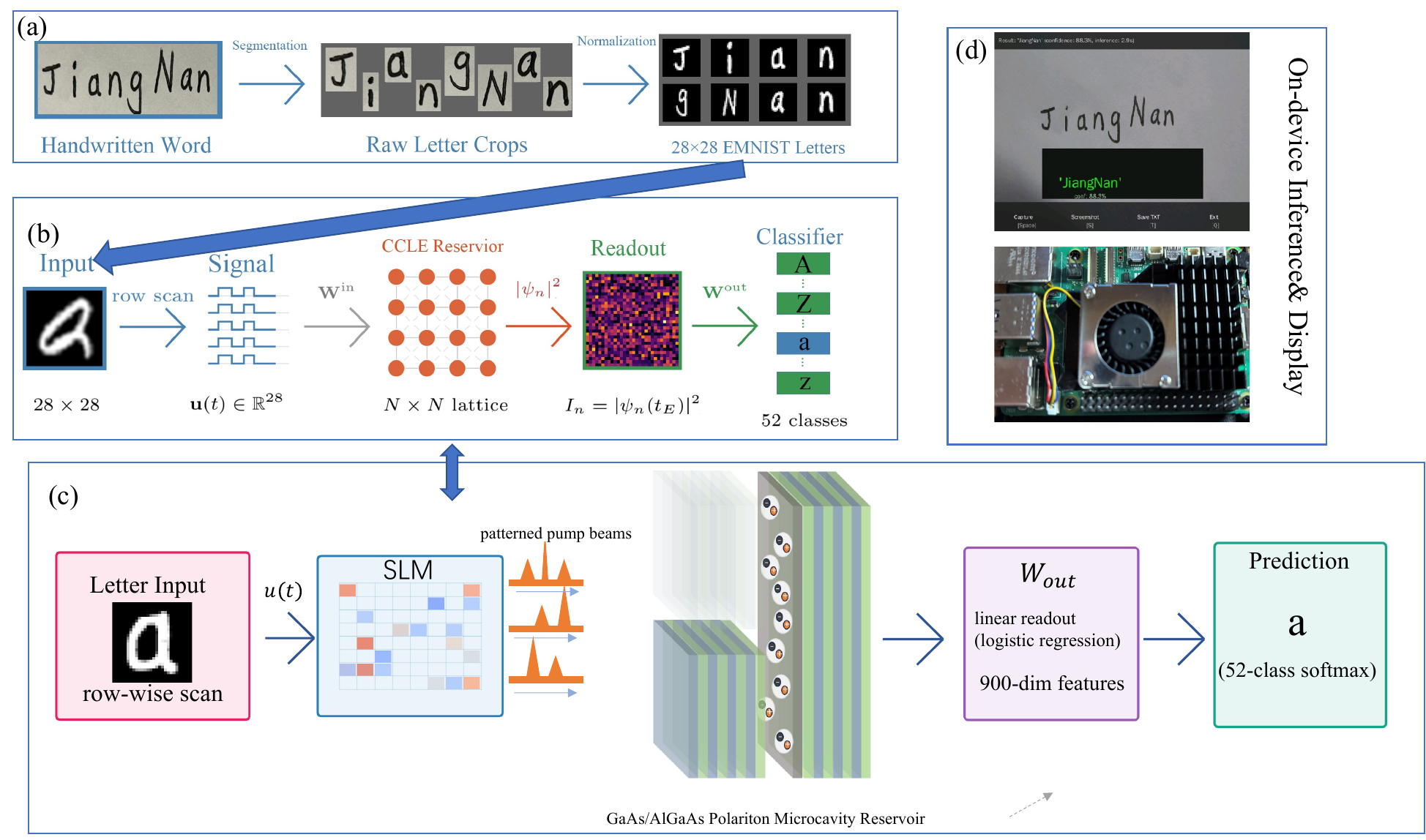}
	\caption{\textbf{Schematic of the full pipeline for handwritten word recognition using CGLE reservoir computing.}
		(a)~Top panel---Segmentation \& Preprocessing: A handwritten word image is first segmented into individual letter crops, which are then normalized to $28\times28$ grayscale images in EMNIST format.
		(b)~Bottom panel---CGLE Reservoir Computing: Each $28\times28$ letter image is processed through five stages.
		(1)~Input---the $28\times28$ grayscale image.
		(2)~Signal---row-wise scanning converts the image into a 28-dimensional temporal signal $\mathbf{u}(t)\in\mathbb{R}^{28}$.
		(3)~CGLE Reservoir---the signal is injected into an $N\times N$ lattice of coupled complex nodes via the random weight matrix $\mathbf{W}^{\rm in}$.
		(4)~Readout---the squared modulus $I_n = |\psi_n(t_E)|^2$ of each node at the end of the input sequence forms a high-dimensional feature vector.
		(5)~Classifier---a linear output layer $\mathbf{W}^{\rm out}$ maps the reservoir state to probabilities over 52 classes (A--Z, a--z).
        (c)~Right---Proposed Physical Implementation: Conceptual pipeline for realizing the CGLE reservoir on a GaAs/AlGaAs exciton-polariton microcavity platform. The six-stage architecture---Letter Input, SLM \& Pump, Polariton Microcavity, CCD Readout, Linear Readout, and 52-Class Prediction---is detailed in the main text. This scheme has not been experimentally realized in this work.
		(d)~Right---On-device Inference \& Display: The trained model deployed on a Raspberry Pi~5 edge device, showing real-time word-level recognition with confidence score.\label{fig:schematic}}
	
\end{figure}

Before deploying our polariton network, we first performed training using the EMNIST dataset \cite{Cohen2017}.
Each handwritten letter image from the EMNIST dataset is a $28 \times 28$ grayscale pixel array.
We convert each image into a spatiotemporal input signal via row-wise scanning, as illustrated in Fig.~\ref{fig:schematic}.
Each of the 28 rows (containing 28~pixels each) is held constant for $\tau_{\rm pixel} = 8$ time steps, producing an input signal $u_m(t)$ of duration $28 \times 8 = 224$ time steps with $D = 28$ input channels.
This encoding converts the spatial structure of each handwritten letter into a temporal sequence that the reservoir can process.

The reservoir dynamics are integrated numerically using the Euler method with time step $dt$.
At the end of the input sequence ($t_E = 224 \cdot dt$), the squared modulus of each node amplitude, $I_n = |\psi_n(t_E)|^2$, is recorded as the reservoir state vector.
This $N^2 = 900$-dimensional vector serves as the feature representation of the input image.

The readout layer computes class scores via a linear transformation
\begin{equation}
\label{eq:readout}
\mathbf{z} = \mathbf{W}^{\rm out} \cdot \mathbf{I},
\end{equation}
\noindent where $\mathbf{W}^{\rm out} \in \mathbb{R}^{52 \times 900}$ and the predicted label is $\hat{c} = \arg\max_c z_c$.
Training minimizes the $L_2$-regularized multinomial cross-entropy loss using the L-BFGS algorithm~\cite{Hastie2009}.
Only the output weight matrix $\mathbf{W}^{\rm out}$ is trained; the reservoir connections $W_{nm}$ and input weights $W_{nm}^{\rm in}$ remain fixed at their random initial values, reducing training to a convex optimization problem with only $52 \times 900 = 46{,}800$ trainable parameters.

We use the EMNIST ``byclass'' dataset, retaining the 52~letter classes (A--Z, a--z) and excluding the 10~digit classes.
Training samples are balanced per class, with 20\% held out for validation.
All experiments use a fixed random seed for reproducibility; test accuracy varies by less than 0.1\% across independent realizations of the random coupling matrices.

\section{Results and Discussion}

\subsection{Overall classification performance}

Unless otherwise noted, the baseline model uses the following parameters: $N = 30$ (900~nodes), nearest-neighbor couplings $W_{nm} \sim \mathcal{U}[0, 0.5]$, next-nearest-neighbor couplings $\sim \mathcal{U}[0, 0.25]$, input weights $W_{nm}^{\rm in} \sim \mathcal{U}[-P/2, P/2]$ with $P = 0.7$, linear gain $\gamma = 0.09$, nonlinear damping $\Gamma = 0.2$, nonlinear coupling $g = 0.1$, and integration time step $dt = 0.03$.

\begin{figure}[htbp]
	\centering
	\includegraphics[width=0.85\textwidth]{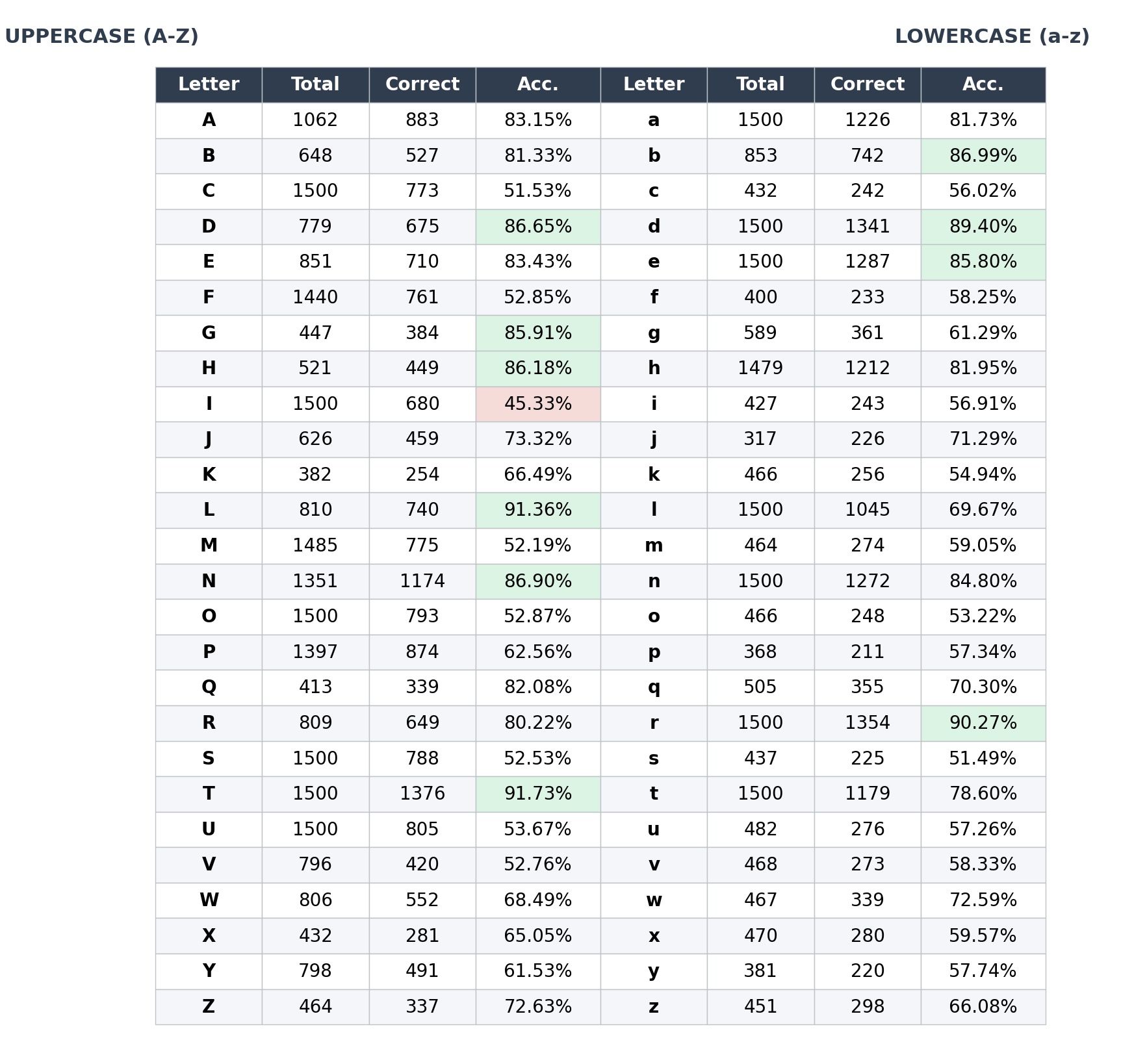}
	\caption{\textbf{Per-letter classification accuracy on the EMNIST 52-class test set.}
		The table shows total samples, correct predictions, and accuracy for each uppercase (A--Z) and lowercase (a--z) letter.
		Cells are color-coded: green highlights indicate high accuracy ($\gtrsim 85\%$), pink indicates notably low accuracy.
		The overall test accuracy is 70.33\%.
		The highest per-letter accuracy is achieved for letters T (91.73\%) and L (91.36\%), while the lowest is I (45.33\%).
		Major confusion occurs between case-sensitive pairs with similar morphology (C/c, I/i, O/o, S/s, U/u, V/v, W/w, X/x, Y/y, Z/z), which is a well-known characteristic of the EMNIST 52-class benchmark. \label{fig:accuracy}}
	
\end{figure}

Fig.~\ref{fig:accuracy} presents the per-letter classification accuracy on the EMNIST~52-class test set.
The system achieves an overall test accuracy of 70.33\%, with training accuracy of 80.41\% and validation accuracy of 69.50\%.
To put this in context, we compare against two baselines.
The chance level for 52-class classification is approximately 1.9\%.
As a linear baseline, we trained logistic regression directly on the 784 raw pixel values (without reservoir), using the same training protocol and regularization.
The pixel-level classifier achieves 61.51\% test accuracy---demonstrating that the raw image features contain substantial class information, but fall considerably short of reservoir-based performance.

The CGLE reservoir therefore provides an 8.8~percentage-point improvement over the linear baseline, showing that the nonlinear high-dimensional embedding performed by the reservoir dynamics yields a meaningful representational advantage.
Furthermore, this is achieved with only 900~reservoir nodes and a linear readout requiring only $900 \times 52 = 46{,}800$ trainable parameters, compared to hundreds of thousands in a typical fully connected network of similar size.

The per-letter accuracy distribution reveals several important features.
Six letters achieve accuracy above 85\%: T (91.73\%), L (91.36\%), r (90.27\%), d (89.40\%), N (86.90\%), and b (86.99\%).
These letters possess distinctive morphological features that are robustly encoded by the reservoir.
Conversely, the worst-performing letters are I (45.33\%), S (52.53\%), s (51.49\%), C (51.53\%), M (52.19\%), and O (52.87\%).
A clear pattern emerges: case-sensitive pairs of visually similar letters---C/c, I/i, O/o, S/s, U/u, V/v, W/w, X/x, Y/y, Z/z---exhibit systematically lower accuracy.
This behavior mirrors the known difficulty of EMNIST 52-class classification, where the case distinction relies on subtle differences in size and proportion rather than shape.

\subsection{Parameter dependence}

\begin{figure}[htbp]
\includegraphics[width=1.0\textwidth]{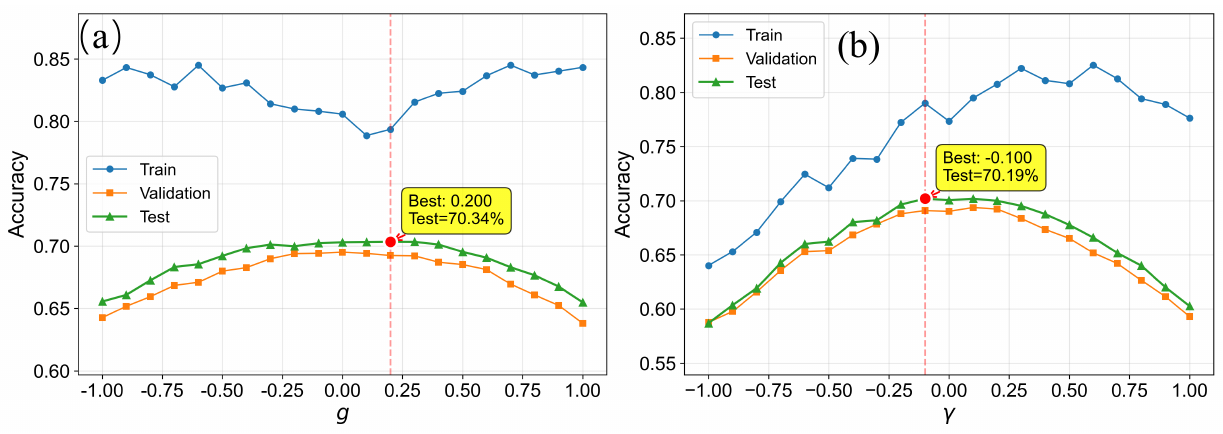}
\centering
\caption{\textbf{Classification accuracy as a function of three key physical parameters.}
(a) Nonlinear coupling $g$: the optimal value is $g = 0.20$ (test accuracy 70.34\%).
(b) Linear gain $\gamma$: the optimal value is $\gamma = -0.10$ (test accuracy 70.19\%).
(c) Time step $dt$: the optimal value is $dt = 0.030$ (test accuracy 70.33\%).
In each panel, blue circles denote training accuracy, orange squares validation accuracy, and green triangles test accuracy.
The pink dashed line marks the optimal value, selected by maximizing test accuracy.
All three parameters exhibit an inverted-U dependence for validation and test accuracy, while training accuracy tends to increase monotonically or remain plateaued, demonstrating a systematic overfitting--generalization trade-off. \label{fig:param_scan}}

\end{figure}

To understand how the physical parameters of the CGLE reservoir influence classification performance, we performed systematic one-dimensional parameter scans.
Fig.~\ref{fig:param_scan} shows the results for three key parameters: the nonlinear coupling strength $g$, and the linear gain $\gamma$.
In each case, the scan reveals an optimum that differs from the default value used for the baseline model, indicating that the defaults are not globally optimal.
Scans for the remaining parameters ($\Gamma$ and $P$) are provided in the Supporting Information.

A common feature across all three parameters is the inverted-U shape of the validation and test accuracy curves, while training accuracy either increases monotonically (for $dt$ and $\gamma$) or oscillates around a high plateau (for $g$).
This divergence between training and generalization performance is a hallmark of overfitting: as the reservoir dynamics become more strongly nonlinear (larger $g$), more dissipative (more negative $\gamma$), or more finely sampled (larger $dt$), the system develops richer internal states that can memorize training examples but fail to generalize to unseen data.
The optimal operating point represents a balance where the reservoir dynamics are sufficiently rich to separate input classes, yet sufficiently regular to maintain the echo state property.


The nonlinear coupling $g$ [Fig.~\ref{fig:param_scan}(a)] is significant because it governs the phase-amplitude interaction that distinguishes CGLE reservoirs from amplitude-only nonlinear systems. The optimal value of $g = 0.20$ places the system in a regime with moderate nonlinear phase modulation.
The relatively flat test accuracy curve for $g \in [0, 0.4]$ suggests that the reservoir is robust to variations in this parameter, which is favorable for physical implementations where precise control of nonlinear coefficients may be challenging. We note that the one-dimensional scans performed here do not capture potential coupling effects between parameters (e.g., the product $g \cdot \gamma$ jointly determining the effective nonlinear gain); a full multi-dimensional optimization remains an interesting direction for future work.

The linear gain $\gamma$ [Fig.~\ref{fig:param_scan}(b)] peaks near $\gamma = -0.10$, slightly below the instability threshold $\gamma = 0$.
This is consistent with the well-established RC design principle that optimal performance occurs when the reservoir operates near the edge of stability~\cite{Lukosevicius2012,Opala2019}.  At $\gamma < -0.2$, the linear damping dominates, and the reservoir states decay toward zero, losing their ability to separate inputs. At $\gamma > 0.1$, the system becomes unstable in the absence of input, and the echo state property degrades.

\subsection{Nonlinear dynamics diagnostics}

\begin{figure}[htbp]
\centering
	\includegraphics[width=0.9\textwidth]{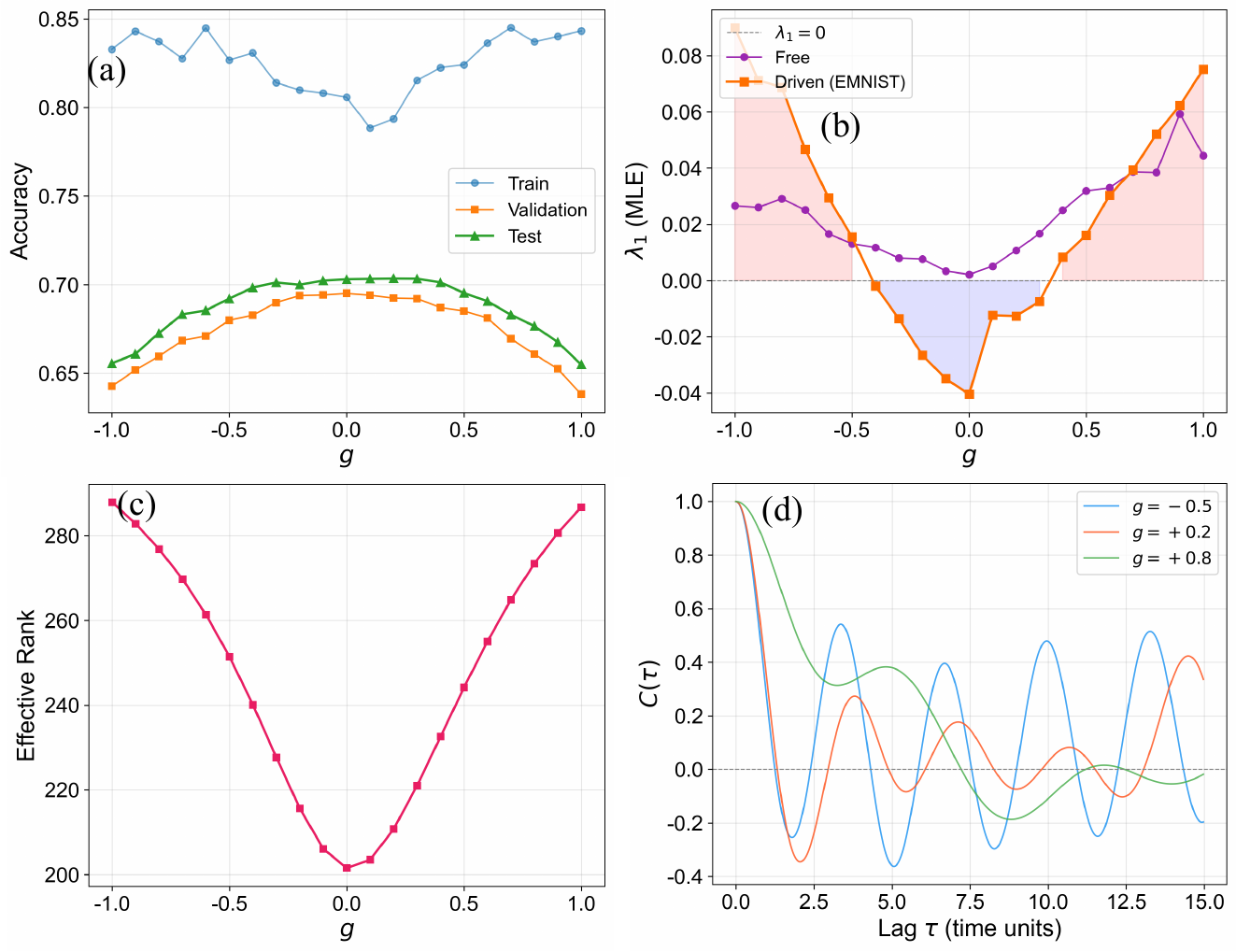}
\caption{\textbf{Nonlinear dynamics of the CGLE reservoir as a function of the nonlinear coupling strength $g$.}
(a) Classification test accuracy vs.\ $g$, showing a peak at $g \approx 0.2$ with accuracy $\sim$70.3\%.
(b) Maximum Lyapunov exponent $\lambda_1$ vs.\ $g$, computed both for the freely evolving reservoir (purple circles) and for the reservoir continuously driven by an EMNIST input sample (orange squares).
The free-evolution MLE remains positive across the full $g$ range, reaching a minimum $\lambda_1 \approx 0.002$ near $g = 0$.
Remarkably, the driven MLE becomes negative ($\lambda_1 < 0$) in the range $g \in [-0.1, 0.4]$, indicating that the input signal stabilizes the reservoir dynamics precisely in the parameter regime where optimal classification occurs.
(c) Effective rank of the reservoir state matrix vs.\ $g$, exhibiting a V-shaped dependence with a minimum of approximately 200 at $g = 0$.
Vertical dashed lines mark the free MLE minimum ($g \approx 0$, purple), the driven MLE zero-crossing ($g \approx -0.1$, orange), and the best classification accuracy ($g \approx 0.2$, green).
(d) Autocorrelation function $C(\tau)$ of freely evolving reservoir node activity for three representative $g$ values.
Smaller $g$ values produce rapid oscillations and fast decorrelation; larger $g$ yields slower, more coherent dynamics with longer memory.}
\label{fig:dynamics}
\end{figure}

To probe the link between reservoir dynamics and computational performance, we analyzed four complementary diagnostics as a function of the nonlinear coupling $g$, summarized in Fig.~\ref{fig:dynamics}: classification accuracy, the maximum Lyapunov exponent (MLE) $\lambda_1$, the effective rank of the reservoir state matrix, and the autocorrelation function $C(\tau)$ of freely evolving node activity.

The MLE quantifies the average exponential rate of divergence (or convergence) of infinitesimally close trajectories, defined as
\begin{equation}
\label{eq:MLE}
\lambda_1 = \lim_{t\to\infty} \lim_{\delta_0 \to 0} \frac{1}{t} \ln \frac{\|\delta\psi(t)\|}{\|\delta\psi_0\|},
\end{equation}
\noindent where $\delta\psi(t)$ is the evolution of an infinitesimal perturbation $\delta\psi_0$ to the system trajectory.
In practice, $\lambda_1$ was computed using the Wolf renormalization method~\cite{Wolf1985,Kantz2003} both for the freely evolving reservoir (no input signal) and for the reservoir continuously driven by a representative EMNIST input sample, thereby probing the stability of the reservoir in its actual operating condition.
In both cases, a 2000-step warmup period was used to ensure convergence to the attractor, and the perturbation was renormalized every 20~time steps over a total of 10,000~steps.

The free-evolution MLE [Fig.~\ref{fig:dynamics}(b), purple curve] reveals a pronounced minimum near $g = 0$, where $\lambda_1^{\rm free} \approx 0.002$, approaching the stability boundary $\lambda_1 = 0$ from above.
As $|g|$ increases, the system becomes increasingly chaotic, with $\lambda_1^{\rm free}$ reaching approximately 0.027 at $g = -1$ and 0.059 at $g = 1$.
The fact that the free MLE remains positive across almost the entire range indicates that the autonomous reservoir operates in a weakly chaotic regime.

The driven MLE [Fig.~\ref{fig:dynamics}(b), orange curve] reveals a qualitatively different picture.
Under continuous input injection, the MLE is substantially reduced across the entire $g$ range and, crucially, becomes negative in the interval $g \in [-0.1, 0.4]$, reaching a minimum of $\lambda_1^{\rm driven} \approx -0.028$ near $g = -0.1$.
A negative MLE indicates that the driven reservoir dynamics are asymptotically stable---neighboring trajectories converge rather than diverge---which is precisely the echo state property required for effective reservoir computing.
The optimal classification accuracy at $g = 0.2$ falls within this driven-stable regime, providing a natural explanation for the apparent offset between the free MLE minimum ($g \approx 0$) and the accuracy peak ($g \approx 0.2$): the input signal itself shifts the effective stability boundary.
We attribute this stabilization to the fact that the input term $W_{nm}^{\rm in} u_m$ in Eq.~(\ref{eq:CGLE}) acts as an effective time-dependent linear gain, which offsets the intrinsic tendency toward chaotic divergence when the nonlinear phase-amplitude coupling $g$ is moderate.

The effective rank $R_{\rm eff}$ [Fig.~\ref{fig:dynamics}(c)] quantifies the dimensionality of the reservoir state space when driven by real input data.
We compute it via the spectral entropy of the normalized singular values:
\begin{equation}
\label{eq:eff_rank}
R_{\rm eff} = \exp\!\left(-\sum_{i} \tilde{\sigma}_i \ln \tilde{\sigma}_i\right), \quad \tilde{\sigma}_i = \frac{\sigma_i}{\sum_j \sigma_j},
\end{equation}
\noindent where $\sigma_i$ are the singular values of the reservoir state matrix $\mathbf{S} \in \mathbb{R}^{M \times N^2}$, constructed from $M = 333$~EMNIST samples over the $N^2 = 900$~reservoir nodes~\cite{7098875,Carroll2019}.
The effective rank shows a clear V-shaped dependence on $g$, decreasing from approximately 287 at $g = \pm 1$ to a minimum of approximately 200 at $g = 0$.
Lower effective rank indicates that the reservoir states occupy a more compact manifold in the high-dimensional space, which reduces the variance of the readout weights and is consistent with improved generalization.
Notably, the effective rank minimum at $g=0$ coincides with the free MLE minimum rather than the accuracy peak at $g=0.2$, suggesting that while low state-space dimensionality favors generalization, a minimal amount of nonlinearity (reflected in the slightly positive driven MLE near the zero boundary) is still required for sufficient class separability.

The autocorrelation function $C(\tau)$ [Fig.~\ref{fig:dynamics}(d)] of freely evolving node activity reveals how the temporal memory of the reservoir depends on $g$.
For $g = -0.5$, the autocorrelation oscillates rapidly with a period of approximately 3~time units and crosses zero multiple times within the first 15~lags, indicating short memory and rich high-frequency content.
For $g = 0.8$, the autocorrelation decays slowly, crossing zero only once at $\tau \approx 6$, indicating longer memory and a smoother dynamical landscape.
The intermediate value $g = 0.2$ (near the optimal classification point) shows moderate decay with balanced oscillatory and persistent components.

Taken together, these diagnostics reveal a consistent dynamical picture.
The reservoir achieves optimal classification in a parameter regime where the input-driven dynamics become marginally stable ($\lambda_1^{\rm driven} \lesssim 0$), the state-space effective rank is low but not minimized, and the temporal correlations balance short-term flexibility with sufficient memory.
This combination of conditions---the \textit{edge of stability under drive}---extends the well-known heuristic that reservoir computers perform best near the stability threshold~\cite{Lukosevicius2012} to a quantitatively characterized, input-dependent criterion.
While the observed correlations between MLE, effective rank, and accuracy do not establish a strictly causal relationship, they provide physically interpretable diagnostics that can guide the principled optimization of CGLE-based and other physical reservoir computers.

\begin{figure}[htbp]
\centering

\includegraphics[width=1\textwidth]{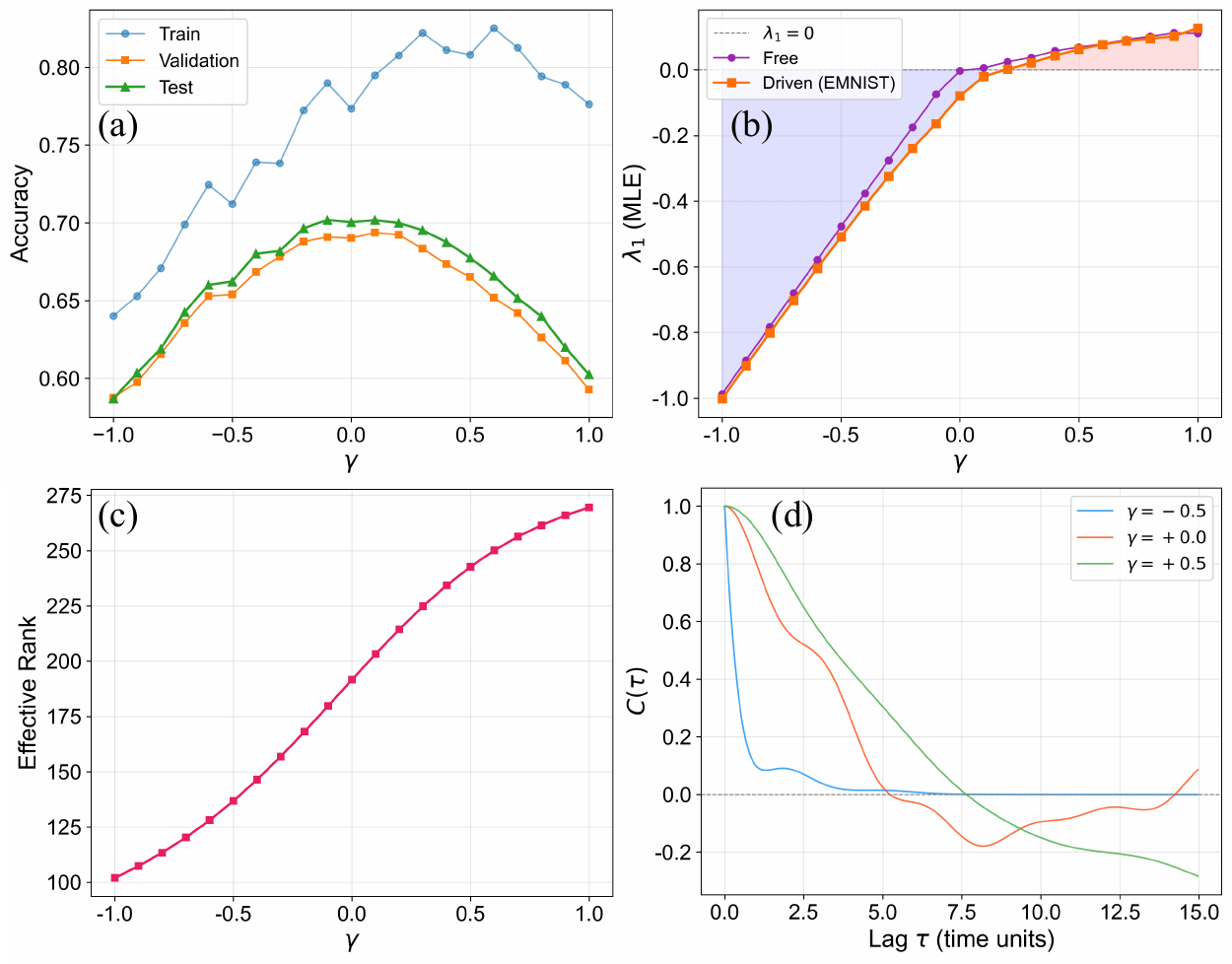}\\

\caption{\textbf{Nonlinear dynamics of the CGLE reservoir as a function of the linear gain $\gamma$.}
(a) Classification test accuracy vs.\ $\gamma$, peaking at $\gamma = -0.10$ (test accuracy 70.19\%).
(b) Maximum Lyapunov exponent $\lambda_1$ vs.\ $\gamma$ for free (purple) and driven (orange) evolution.
In contrast to the V-shaped $g$-dependence, both MLE curves increase monotonically with $\gamma$, crossing the stability boundary $\lambda_1 = 0$ near $\gamma \approx 0$.
The optimal classification accuracy at $\gamma = -0.10$ corresponds to $\lambda_1^{\rm free} \approx -0.075$---the system operates precisely at the edge of chaos on the stable side.
(c) Effective rank vs.\ $\gamma$, also exhibiting a monotonic increase, from $\sim$104 at $\gamma = -1.0$ to $\sim$270 at $\gamma = +1.0$.
The working point $\gamma = -0.10$ yields $R_{\rm eff} \approx 180$, providing sufficient state-space dimensionality without the excessive noise associated with strongly chaotic dynamics.
(d) Autocorrelation function $C(\tau)$ for three representative $\gamma$ values.
At $\gamma = -0.5$ (deeply stable), the autocorrelation decays slowly, indicating long memory but sluggish response to input variations.
At $\gamma = +0.5$ (chaotic), $C(\tau)$ collapses to zero almost immediately, signifying loss of fading memory.
At $\gamma \approx 0$ (near the stability boundary), the decay is moderate, balancing memory retention with dynamic responsiveness.}
\label{fig:dynamics_gamma}
\end{figure}

To further probe the generality of the edge-of-chaos principle, we performed the same set of dynamical diagnostics as a function of the linear gain $\gamma$, summarized in Fig.~\ref{fig:dynamics_gamma}.
Whereas the nonlinear coupling $g$ governs the phase-amplitude interaction, the linear gain $\gamma$ directly controls the net amplification of the field: negative $\gamma$ damps perturbations toward the trivial fixed point $\psi_n = 0$, while positive $\gamma$ drives exponential growth that is eventually saturated by the nonlinear damping term $\Gamma|\psi_n|^2$.
This fundamental difference in physical mechanism leads to qualitatively distinct dynamical signatures.

The free-evolution MLE [Fig.~\ref{fig:dynamics_gamma}(b), purple curve] exhibits a monotonic dependence on $\gamma$, in striking contrast to the V-shaped $g$-dependence.
As $\gamma$ increases from $-1.0$ to $+1.0$, $\lambda_1^{\rm free}$ rises from approximately $-0.99$ (strongly stable) to $+0.11$ (chaotic), crossing the stability boundary $\lambda_1 = 0$ at $\gamma \approx 0$.
The optimal classification accuracy at $\gamma = -0.10$ corresponds to $\lambda_1^{\rm free} \approx -0.075$---the system operates barely on the stable side, precisely at the \textit{edge of chaos}.

The driven MLE [Fig.~\ref{fig:dynamics_gamma}(b), orange curve] follows a parallel monotonic trend, remaining systematically lower than the free MLE across the entire $\gamma$ range due to the stabilizing effect of the input signal.
At the optimal working point $\gamma = -0.10$, the driven MLE is also negative, confirming that the reservoir satisfies the echo state property under actual operating conditions.

The effective rank [Fig.~\ref{fig:dynamics_gamma}(c)] mirrors the monotonic behavior of the MLE, growing from $R_{\rm eff} \approx 104$ at $\gamma = -1.0$ to $R_{\rm eff} \approx 270$ at $\gamma = +1.0$.
This reflects the increasingly complex state-space structure enabled by stronger linear gain: at large negative $\gamma$, the dynamics collapse toward the trivial fixed point, producing low-dimensional states; at large positive $\gamma$, chaotic divergence generates high-dimensional, noisy representations.
The optimal working point $\gamma = -0.10$ yields $R_{\rm eff} \approx 180$, an intermediate dimensionality that provides sufficient separability between classes while avoiding the excessive variance associated with strongly chaotic dynamics.

The autocorrelation function $C(\tau)$ [Fig.~\ref{fig:dynamics_gamma}(d)] of freely evolving node activity reveals how temporal memory varies with $\gamma$.
At $\gamma = -0.5$ (deeply stable), $C(\tau)$ exhibits slow, oscillatory decay, indicating long intrinsic memory but sluggish response to input variations---the reservoir is too ``sticky.''
At $\gamma = +0.5$ (chaotic), $C(\tau)$ collapses to near zero within a few time units, meaning the reservoir forgets past inputs almost instantly and cannot integrate temporal information over the 224-step input sequence.
At $\gamma \approx 0$ (near the stability boundary), the decay is moderate, balancing memory retention with dynamic responsiveness---precisely the fading memory property required for effective temporal processing.

Taken together, the $\gamma$-dynamics diagnostics reinforce the edge-of-chaos picture from a complementary direction.
The reservoir achieves optimal classification where the free MLE is closest to zero from the stable side, the effective rank is at an intermediate level, and temporal correlations exhibit balanced decay.
While the $g$-dependence reveals optimal performance where the driven system is most stable (free MLE positive, driven MLE negative), the $\gamma$-dependence reveals optimal performance at the transition from stable to chaotic dynamics (both free and driven MLE negative).
In both cases, the optimal operating point lies near the stability boundary, confirming that marginal stability under input drive is a universal design principle for CGLE reservoir computers.

\section{Summary and Outlook}

In summary, we have demonstrated a comprehensive scheme for exciton-polariton reservoir computing using a discrete CGLE lattice. By applying it to the 52-class EMNIST dataset, we show that the system achieves robust classification performance (70.33\% accuracy) with minimal training overhead. Our key contribution lies in the rigorous characterization of the parameter space: we quantitatively show that optimal performance universally occurs at the edge of stability, where the driven system's maximum Lyapunov exponent approaches zero from the negative side. This finding transcends specific parameter choices and provides a fundamental design rule for physical RC systems.

Furthermore, the successful deployment on a Raspberry Pi 5 validates the lightweight nature of the trained model, paving the way for low-power edge computing applications. Looking ahead, while this work focuses on numerical simulation of the CGLE, the ultimate goal is physical realization using patterned microcavity arrays or perovskite lattices. In such systems, the intrinsic picosecond-scale dynamics of polaritons could replace numerical integration, potentially enabling ultrafast, all-optical inference at speeds unattainable by conventional electronics.

\section*{Acknowledgements}

The authors thank  the National Natural Science Foundation of China (Grant No. 12404362) and the Fundamental Research Funds for the Central Universities (Grant No. JUSRP123027).

\section*{Competing interests:}
There are no competing interests to declare.

\section*{Data and materials availability:}
All source code, trained model weights, and numerical data supporting this study are available at Zenodo \url{https://doi.org/10.5281/zenodo.21158082}.

\bibliography{paper.bib}




  

\end{document}